\documentclass[aps,prl,reprint,groupedaddress, onecolumn]{revtex4-2}
\usepackage{mathrsfs}
\usepackage{cancel}
\usepackage{amssymb}
\usepackage{comment}
\usepackage{bm}
\usepackage{amsmath}

\usepackage{graphicx}

\def\p{\partial}  
  
\def\o{\over}
\def\p{\partial}
\def\l{\left}
\def\r{\right}

\def\U5{\tilde U_5}
\def\be{\begin{equation}}
\def\ee{\end{equation}}
\def\bea{\begin{eqnarray}}
\def\eea{\end{eqnarray}}

\begin{document}

%\preprint{}
\parskip=3pt

%Cosmic gravitomagnetic forces in boosted inertial frames \\ 
\title{Inductive rectilinear frame dragging and\\
 local coupling to the gravitational field of the universe}
%added Nathan author 3 May. Discovered Einstein Hamiltonian not a scalar
%added 8 May added Nathan insight about inductive, not gravitomagnetic, dragging

\author{L.L.~Williams}
\email{willi@konfluence.org}

%\homepage[www.konfluence.org]{}
%\thanks{}
\affiliation{Konfluence Research Institute,\\ Manitou Springs, Colorado}

\author{N.~Inan}
\email{ninan@ucmerced.edu}
\affiliation{School of Natural Sciences\\University of California, Merced\\Merced, California}

\date{11 July 2021}
\begin{abstract}
There is a drag force on objects moving in the background cosmological metric, known from galaxy cluster dynamics. The force is quite small over laboratory timescales, yet it applies in principle to all moving bodies in the universe. It means it is possible for matter to exchange momentum and energy with the gravitational field of the universe, and that the cosmological metric can be determined in principle from local measurements on moving bodies. The drag force can be understood as inductive rectilinear frame dragging. This dragging force exists in the rest frame of a moving object, and arises from the off-diagonal components induced in the boosted-frame metric. Unlike the Kerr metric or other typical frame-dragging geometries, cosmological inductive dragging occurs at uniform velocity, along the direction of motion, and dissipates energy. Proposed gravito-magnetic invariants formed from contractions of the Riemann tensor do not appear to capture inductive dragging effects, and this might be the first identification of inductive rectilinear dragging.
\end{abstract}
\maketitle
%%%%%%%%%%%%%%%%%%%%%%%%%%%%%%%%%%%%%%%%%%%%%
%%%%%%%%%%%%%%%%%%%%%%%%%%%%%%%%%%%%%%%%%%
\section{1. Introduction}
%%%%%%%%%%%%%%%%%%%%%%%%%%%%%%%%%%%%%%%%%%%%%
%%%%%%%%%%%%%%%%%%%%%%%%%%%%%%%%%%%%%%%%%%
The freedom of a general coordinate transformation allows a local Lorentz frame to be defined sufficiently locally around a point in curved spacetime. In this frame, also called an inertial frame, first derivatives of the metric vanish, and therefore the connections vanish, and gravitational forces on moving bodies vanish. Yet the coordinate freedom cannot remove all second derivatives of the metric, and so components of the Riemann tensor can be non-zero locally where the connections are zero. The lengthscale of the second derivatives determines the scale of the spacetime region over which first derivatives vanish and the local Lorentz frame holds.\cite{wbg},\cite{pw}

This local vanishing of gravitational forces, and the resulting local rectilinear trajectories of moving bodies in these frames, can be understood as an expression of the Equivalence Principle. Gravitational forces can be made to vanish locally. Correspondingly, it is impossible to localize gravitational field energy.\cite{mtw}

Now let us ask whether it is possible for matter to {\it locally} exchange momentum or energy with the gravitational field of the universe. According to a naive understanding of the Equivalence Principle, and the availability of local Lorentz frames, it might seem impossible for a body to exchange momentum or energy with the gravitational field of the universe, if such energy cannot be localized in the field. This article is to show how such exchange can in principle occur locally.

The gravitational field of the universe is the cosmological metric of general relativity. Since the discovery of dark energy at the turn of the century, the cosmological metric is well-constrained to behave as a flat Robertson-Walker metric, and the energy budgets of matter, radiation, and dark energy are well-constrained. The resulting cosmological model is called Lambda-Cold-Dark-Matter, where Lambda refers to dark energy.\cite{fth}

For bodies in motion against the cosmological metric, there is a drag force resulting from the expanding spacetime.\cite{crl},\cite{isl} This force is sometimes called Hubble drag, e.g.\cite{pck},\cite{ldr} and has been in models of galaxy cluster dynamics for decades\cite{P1},\cite{P2}. 

%The mathematics that leads to this relativistic force hold also for small bodies and even mass points. As such, this force qualifies as a direct local coupling between a moving body and the gravitational field of the universe.

We show that the Hubble drag force exists in the rest frame of a moving body. Although gravitational forces are coordinate dependent, the Hubble drag force exists in all frames, and there is an invariant expression for the %deflection of a hypothetical spring gauge measuring the 
force.

We propose that Hubble drag can be understood as inductive rectilinear frame dragging. Rectilinear, or translational, frame-dragging has been considered to some extent for linear acceleration.\cite{ge},\cite{lb},\cite{fz},\cite{pk},\cite{pfr},\cite{pfs} However, Hubble drag would be the first case of rectilinear frame dragging proposed for an unaccelerated body.

Hubble drag can be understood as a type of frame-dragging because it arises from off-diagnonal metric components, as in the Kerr metric around rotating bodies. Yet in the boosted frame, Hubble drag originates not from the gravito-magnetic field, but from the inductive part of the gravito-electric field. The gravito-magnetic field vanishes in the cosmological case because there are no spatial curls, and the Newtonian part of the gravito-electric field vanishes for the same reason. Therefore proposed gravito-magnetic invariants formed from contractions of the Riemann tensor \cite{cfi1} do not capture inductive gravito-electric effects.

%%%%%%%%%%%%%%%%%%%%%%%%%%%%%%%%%%%%%%%%%%%%%
%%%%%%%%%%%%%%%%%%%%%%%%%%%%%%%%%%%%%%%%%%
\section{2. Force on a body moving in the isotropic frame}
%%%%%%%%%%%%%%%%%%%%%%%%%%%%%%%%%%%%%%%%%%%%%
%%%%%%%%%%%%%%%%%%%%%%%%%%%%%%%%%%%%%%%%%%

The standard model of cosmology, the Lambda-Cold-Dark-Matter model, describes the metric of the universe in terms of the Robertson-Walker metric, with zero curvature \cite{fth}:
\be
\label{rwm}
-c^2 d\tau^2 = -c^2 dt^2 + a^2(t) [dx^2 + dy^2 + dz^2]
\ee
in terms of cosmological time coordinate $t$ and spatial coordinates $x,y,z$, where $c$ is the speed of light, and $a(t)$ is the cosmological scale factor. The scale factor at the present epoch $t_0$ is $a(t_0)=1$. 

The Hubble constant is given in terms of the scale factor by 
\be
\label{hub}
H(t) \equiv {1 \o a}{da\o dt} 
\ee
The Hubble constant at the current epoch is $H(t_0) \equiv H_0 = (da/dt)\vert_{t_0}$, the cosmological value observed today.

The metric (\ref{rwm}) is the gravitational field of the universe. It has the form of being maximally symmetric in the spatial components, and independent of position. This form enforces the standard assumption of an isotropic and homogeneous universe. The metric (\ref{rwm}) is characterized by the single parameter $a(t)$, where $t$ is a cosmological time coordinate that goes to $t=0$ at the Big Bang. The exact functional dependence of $a(t)$ depends on the energy content of the universe, with distinct regimes for radiation-dominated, matter-dominated, and Lambda-dominated phases of evolution.

The form (\ref{rwm}) can be considered a cosmic standard coordinate system, and the time coordinate a cosmic standard time coordinate. The time coordinate can be determined in principle from local time-dependent cosmological measurements, such as the temperature change of the microwave background, or the acceleration/deceleration of the redshift.\cite{wbg2} The time coordinate of (\ref{rwm}) is the time measured by a local free-fall observer.

The motion of a free body in spacetime is described by the geodesic equation,
\be
\label{ge}
{dU^\mu\o d\tau} + \Gamma^\mu_{\alpha\beta} U^\alpha U^\beta = 0
\ee
where $c^2d\tau^2 = g_{\mu\nu} dx^\mu dx^\nu$, greek indices range over the 4 components of spacetime, and the 4-velocity of a body is
\be
U^\mu = {dx^\mu\o d\tau}
\ee
The connections $\Gamma^\mu_{\alpha\beta}$ are given in terms of the metric by the standard textbook formula. From a cosmological perspective, the relativistic force equation (\ref{ge}) is understood to apply to galaxies or clusters of galaxies, moving in the smoothed, cosmological metric (\ref{rwm}). Local gravitational interactions will always dominate cosmic gravitational influences.

For a body at rest in the cosmic rest frame of (\ref{rwm})
\be
{\widetilde U}^\mu = (c, 0,0,0)
\ee
The force equation is simply
\be
\label{ger}
{d{\widetilde U}^\mu\o d\tau} + {\Gamma}^\mu_{tt} c^2 = 0
\ee
The connections operative for a body at rest in (\ref{rwm}) are:
\be
\label{rff}
{\Gamma}^\mu_{tt} = {1\over 2} g^{\mu t} \p_t g_{tt} + g^{\mu k} \p_t g_{tk} = 0
\ee
Galaxies at rest in this frame are in free fall. They experience no forces and remain at rest. The time and space coordinates are co-moving.

However, the free-fall frames in which the metric has the form (\ref{rwm}) are not inertial frames, because the connections do not all vanish. The definition of an inertial frame is that {\it all} first derivatives of the metric vanish, even while second derivatives may be present. Therefore, all gravitational forces must vanish in true inertial frames. Yet in the cosmic rest frame (\ref{rwm}), first derivatives do not all vanish. In fact, first derivatives correspond to the Hubble constant. Therefore, a true inertial cosmic frame would stipulate zero Hubble constant, locally. 

Let us consider the connections for the metric (\ref{rwm}). We will use small roman indices $i,j,k$ to denote spatial coordinates, and the index $t$ for the time coordinate. The metric (\ref{rwm}) components are 
\be
g_{tt}=-c^2 \quad,\quad g_{ij}=a^2(t)\ \delta_{ij} \quad,\quad g^{tt} = -c^{-2} \quad,\quad g^{ij} = a^{-2}(t)\ \delta^{ij}
\ee 
Then the connections are:
\be
\label{c1}
\Gamma^i_{tt} = 0 \quad,\quad \Gamma^l_{ij} = 0 \quad,\quad \Gamma^t_{jt} = 0
\quad,\quad \Gamma^t_{tt} =0
\ee
\be
\label{c2}
\Gamma^i_{jt}= {1\o 2} g^{ik} ( {\p_t g_{jk}} + {\cancelto{0}{\p_kg_{tk}} - \cancelto{0}{\p_k g_{tj}}} ) = H \delta^i_j
 %\quad,\quad \delta^j_i \Gamma^i_{jt} =  3 H
\ee
\be
\label{c3}
\Gamma^t_{ij}={1\o 2} g^{tt} ( \cancelto{0}{\p_i g_{jt}} + \cancelto{0}{\p_j g_{it}} - {\p_t g_{ij}} )  = {a^2\o c^2} H\delta_{ij} %\quad,\quad \delta^{ij} \Gamma^t_{ij} = 3{a^2\o c^2} H
\ee
The non-zero spatial connections have 1 time index and are proportional to the Hubble constant (\ref{hub}). They originate in the time dependence of the spatial components $g_{jk}(t)$.

Consider the forces on a body moving in the x-direction with speed $dx/dt \equiv v$ and with 4-velocity
\be
\label{4vel}
U^\mu = (U^t, U^x, 0, 0) = \l( c{dt\o d\tau}, {dx\o d\tau}, 0, 0 \r) = {dt\o d\tau}(c, v,0,0)
\ee

The energy and momentum effects from the cosmological metric are:
\be
{dU^t\o d\tau} + \Gamma^t_{xx} (U^x)^2 = 0
\ee
\be
{dU^x\o d\tau} + 2\Gamma^x_{tx} U^t U^x = 0
\ee

Using the connections (\ref{c1}), ({\ref{c2}), (\ref{c3}), these equations reduce to
\be{
\label{hc}
{dU^t\o dt} = - a^2 H U^t {v^2\o c^2}
}\ee
\be{
\label{hd}
{dU^x\o dt} = - 2H U^x
}\ee
The equation (\ref{hd}) masks a cubic velocity dependence which can be exposed by expanding (\ref{hd}) and using (\ref{hc}):
\be
\label{hdu}
{dv\o dt} = -2Hv + {a^2 v^3\o c^2} H = -Hv\l(2-{a^2v^2\o c^2}\r) = -Hv\l(1+ {c^2\o (U^t)^2}\r) < 0
\ee
where the last equality follows because
\be
-c^2 = g_{\mu\nu} U^\mu U^\nu = -(U^t)^2 + a^2 (U^x)^2 = -(U^t)^2 (1 - a^2 v^2/c^2)
%\quad\rightarrow\quad U^t = c(1 - a^2 v^2/c^2)^{-1/2}
\ee
which is in turn consistent with (\ref{hc}). This shows that the coordinate acceleration is negative-definite.

%\section{3. Alternate calculation from a conserved quantity}

We can also calculate this force from an alternative form of the geodesic equation that is useful when the metric is independent of a coordinate. In this form, the change to each component of the covariant 4-velocity depends on the derivative of the metric with respect to the corresponding coordinate:
\be
\label{cge}
{dU_\mu\o d\tau} = {1\o 2} U^\alpha U^\beta \partial_\mu g_{\alpha\beta}
\ee

Since the metric (\ref{rwm}) is independent of position, the evolution of spatial velocity of a body is given simply from (\ref{cge}) by:
\be
U_j = g_{j\mu} U^\mu = a^2 U^x = \text{constant}
\ee
which implies:
\be
\label{hd2}{
 {dU^x\o dt} = -  2 H U^x 
}\ee
This is the same cosmological drag equation (\ref{hd}). Although it is written for a particular component here, the isotropy of the Hubble expansion means it will be experienced by an object with any rectilinear velocity. This allows us to understand Hubble drag in terms of conservation of momentum in the expanding universe.

%\begin{comment}
%\section{4. Associated geodesic deviation calculation}

The Hubble drag force has an associated effect in the geodesic deviation equation
\be
{d^2 S^\mu\o d\tau^2} = R^\mu_{\nu\rho\sigma} U^\nu U^\rho S^\sigma
\ee
where $U^\nu$ is the 4-velocity of a moving body, and $S^\nu$ is a deviation vector separating observers with that 4-velocity.

The independent components of the Riemann tensor are
\be
c^2 R^k_{ttj} = {\bm\ddot a\o a} \delta^k_j \quad,\quad c^2 R^i_{jkl} = a^2 H^2 (\delta^i_k \delta_{jl} - \delta^i_l \delta_{jk} )
\ee
and their non-zero permutations, which may differ by factors of $a^2$.

For the case (\ref{4vel}) of a body moving in the $x$ direction, there will be geodesic deviation along $x$:
\bea
\label{gde}
{d^2 S^x\o d\tau^2} &=& (R^x_{tt\mu} U^t U^t + {R^x_{tx\mu}} U^t U^x  + \cancelto{0}{R^x_{xt\mu}} U^x U^t + \cancelto{0}{R^x_{xx\mu}} U^x U^x )S^\mu /c^2 \nonumber \\ &=& R^x_{ttx} U^t U^t S^x/c^2 + {R^x_{txt}} U^t U^x S^t /c^2
= {\bm\ddot a\o a} U^t (U^t S^x - U^x S^t)/c^2 \ne 0
\eea

This shows that the Hubble drag force corresponds to a non-vanishing effect in geodesic deviation, along the direction of motion of moving bodies. There is a well-known increasing spatial spatial separation of bodies at rest accruing from the expansion of the universe. For moving bodies there is also acceleration along the direction of motion.
%\end{comment}

%\subsection{5. Hubble drag as a local gravitational phenomenon}

%The equation (\ref{hc}) represents the reduction in kinetic energy that accrues from cosmic expansion. 

The equation (\ref{hd}) expresses the well-known effect from galaxy cluster dynamics \cite{P1},\cite{P2} that is sometimes known as Hubble drag, e.g.\cite{pck},\cite{ldr}. It implies that an observer at rest in the cosmic standard coordinate system would detect a slowing and a drag force on a moving body. The result (\ref{hd2}) from a conserved quantity is also well known.\cite{crl},\cite{isl}

The force is minute, as the characteristic time scale for this slow-down is the age of the universe:
\be
\label{sz}
\Delta U \sim {\Delta t} H U \sim {\text{dynamical timescale}\o \text{age of universe}}\ U
\ee
 This Hubble drag force would be undetectable for terrestrial or planetary phenomena. It is relevant for super clusters or other objects that are in cosmic free fall over the age of the universe.

The Hubble drag force effect is usually understood to apply to galaxy clusters,\cite{P1,P2} but mathematically it applies at a point in the smoothed background metric. There is a non-zero effect in the geodesic deviation (\ref{gde}) that validates the reality of the local gravitational interaction. Therefore these force effects should be understood as local gravitational phenomena arising from an interaction with the gravitational field of the universe. 

Here is the interesting case of action of the gravitational field of the universe at a point, and on a body in uniform motion. It seems to imply that a local experiment on a small moving body deep in intergalactic space can in principle measure the gravitational field of the universe.

%%%%%%%%%%%%%%%%%%%%%%%%%%%%%%%%%%%%%%%%%%%%%
%%%%%%%%%%%%%%%%%%%%%%%%%%%%%%%%%%%%%%%%%%
\section{3. Force in the rest frame of a moving body}
%%%%%%%%%%%%%%%%%%%%%%%%%%%%%%%%%%%%%%%%%%%%%
%%%%%%%%%%%%%%%%%%%%%%%%%%%%%%%%%%%%%%%%%%
Now we show that the Hubble drag force manifests in the rest frame of a moving body, and therefore constitutes an invariant rest-frame force.

The isotropy of the cosmological metric (\ref{rwm}) selects a preferred cosmological frame. There is only one frame in which the metric is isotropic. Any coordinate transformation involving a boost at constant velocity will introduce off-diagonal components into the transformed isotropic metric. For a coordinate system not at rest in (\ref{rwm}), we can write the transformed metric as ${\widetilde g}_{\mu\nu}$, and the components of the connections that operate on a body at rest (\ref{rff}) in this frame are rewritten for ${\widetilde g}_{\mu\nu}$:
\be
\label{rffg}
{\widetilde\Gamma}^\mu_{tt} = {1\over 2} {\widetilde g}^{\mu t} \p_t {\widetilde g}_{tt} + {\widetilde g}^{\mu k} \p_t {\widetilde g}_{tk} \ne 0
\ee

In order to define the boosted frame metric ${\widetilde g}_{\mu\nu}$ from the isotropic metric (\ref{rwm}), we want to consider a velocity transformation with magnitude $v$ in the x-direction. It is clear that the spatial coordinates will be boosted by the velocity, but the time coordinate can be handled in several ways. 

\begin{enumerate}
\item{}Boost to an inertial frame, where all first derivatives vanish, including the Hubble constant
\be
{\widetilde t} = p(x^\mu, v) \quad ,\quad {\widetilde x}^i = q^i(x^\mu, v)
\ee
where $p$ and $q^i$ are functions of the coordinates that eliminate all first derivatives of the metric.
\item{}Lorentz transformation to a boosted frame
\be
\label{t2}
c{\widetilde t} = \gamma (ct + vx/c) \quad,\quad {\widetilde x}= \gamma (x + vt) \quad,\quad 
{\widetilde y} = y \quad, \quad {\widetilde z} =z
\ee
\item{}Galiean-type transformation to a boosted frame, retaining the cosmic standard time coordinate
\be
\label{t3}
{\widetilde t} = t \quad,\quad {\widetilde x}= x + vt \quad,\quad 
{\widetilde y} = y \quad, \quad {\widetilde z} =z
\ee
\end{enumerate}
Although the transformation (\ref{t3}) has the form of a Galilean transformation, and although a Galilean transformation approximates a Lorentz transformation (\ref{t2}), the transformation (\ref{t3}) is valid and exact for relativistic velocities in curved space. The transformation (\ref{t3}) only approximates (\ref{t2}) against a background Minkowski spacetime. In our case, the reference frame of (\ref{rwm}) is not an inertial frame. In order for a boosted object to maintain the cosmic time coordinate in curved space, the transformed metric contains the information for the time dilation that would ordinarily be associated with the transformation (\ref{t2}). So the significance of the choice (\ref{t3}) is not as a non-relativistic approximation to (\ref{t2}), but as the cosmic time coordinate in a boosted frame.

%(aftr 29)  (after 36) Although (29) seems to imply that terms quadratic in v must be neglected, in actuality the metric may still contains such terms which are necessary to preserve the cosmological time coordinate even for a relativistic boost.

Nonetheless, in practice, the 3 velocity transformation cases become indistinguishable for terrestrial time scales and non-relativistic velocities. The first derivatives of the isotropic metric (\ref{rwm}) are quite small -- time derivatives are of order the inverse of the age of the universe -- and so the deviations from Minkowski space are all small. Therefore the cosmic rest frame, with a non-zero, local Hubble constant, is approximately inertial to the extent the Hubble constant can be set to zero. 
%On timescales much less than the age of the universe, the correct velocity transformation is a Lorentz transformation. And for velocities much less than the speed of light, the Lorentz transformation approximates a Galilean transformation. Therefore a non-relativistic boost will preserve the cosmic standard time coordinate.
%LW here is where an explanation is needed about non-relativistc vs cosmic std. Check especially last sentence. Rmove approximates

%Putting these together, let us consider dynamics on timescales much less than the age of the universe, while keeping first derivatives of the metric, and 
Let us therefore choose the transformation (\ref{t3}) that preserves the cosmic standard time coordinate and that approximates (\ref{t2}) when the metric is Minkowski. 
%%c{\widetilde t} = ct \quad,\quad {\widetilde x}= x + vt \quad,\quad 
%{\widetilde y} = y \quad, \quad {\widetilde z} =z
%\ee
%NI: is his redundanct wtih 29?

The components of the transformation matrix for (\ref{t3}) are:
\be
{\p{\widetilde t}\o \p t} = 1 \quad,\quad {c\p{\widetilde t}\o \p x} = 0 \quad,\quad
{\p{\widetilde x}\o \p t} = v \quad,\quad {\p{\widetilde x}\o \p x} = 1 \quad,\quad
{\p{\widetilde y}\o \p y} = 1 \quad,\quad {\p{\widetilde z}\o \p z} = 1
\ee

The inverse transformation matrix components are:
\be
t = {\widetilde t}  \quad,\quad x= {\widetilde x} - v{\widetilde t} \quad,\quad 
y={\widetilde y}  \quad, \quad z={\widetilde z}
\ee
The components of the inverse transformation are:
\be
{\p t\o \p {\widetilde t}} =1 \quad,\quad {c\p t\o \p {\widetilde x}}= 0\quad,\quad
{\p x\o \p {\widetilde t}} =-v \quad,\quad {\p x\o \p {\widetilde x}} = 1 \quad,\quad
{\p y\o \p {\widetilde y}} = 1 \quad,\quad {\p z\o \p {\widetilde z}} =1
\ee
%note to Nathan: the minus sign comes from the inverse transf., not from a direction choice

The transformed metric in the boosted frame is given by
\be
{\widetilde g}_{\mu\nu} = {\p x^\alpha\o\p {\widetilde x}^\mu}{\p x^\beta\o\p {\widetilde x}^\nu} g_{\alpha\beta} = - c^2 {\p t\o\p {\widetilde x}^\mu}{\p t\o\p {\widetilde x}^\nu}  
+ a^2 \delta_{ij} {\p x^i\o\p {\widetilde x}^\mu}{\p x^j\o\p {\widetilde x}^\nu} 
\ee

The components of the metric implicated in the rest-frame forces of (\ref{rffg}) are given by:
\be
{\widetilde g}_{tt} = -1 + {a^2 v^2\o c^2} \quad ,\quad 
{\widetilde g}_{xx}  = a^{2} 
\ee
\be
\label{gm}
{\widetilde g}_{tx} =  -{v\o c}a^2
\ee
The off-diagonal, gravito-magnetic metric components (\ref{gm}) depend on the non-Minkowski character of the isotropic metric (\ref{rwm}). For a truly Minkowskian metric, with no first derivatives, then there are no off-diagonal components because the Minkowski metric is invariant under a Lorentz transformation.
%NI add comment about boost direction

The inverse transformed metric components are given by
\be
{\widetilde g}^{\mu\nu} = {\p {\widetilde x}^\mu\o\p x^\alpha }{\p {\widetilde x}^\nu\o\p x^\beta } g^{\alpha\beta} = 
- {1\o c^4}{\p {\widetilde x}^\mu\o\p t }{\p {\widetilde x}^\nu\o\p t }  
+ a^{-2} \delta_{ij} {\p {\widetilde x}^\mu\o\p x^i }{\p {\widetilde x}^\nu\o\p x^j }
\ee

The components of the inverse metric relevant to rest-frame forces are
\be
{\widetilde g}^{tt} = -c^{-2} \quad ,\quad {\widetilde g}^{xx}  = a^{-2} - {v^2\o c^2}
\ee
%this eqn is ok, but Nathan said there was a sign error
\be
{\widetilde g}^{tx} = - {v\o c}
\ee

Therefore the rest frame connections (\ref{rffg}) are
\be
{\widetilde \Gamma}^t_{tt} ={1\o 2} {\widetilde g}^{tt} {\p_t {\widetilde g}_{tt}} 
+ {\widetilde g}^{tk} {\p_t {\widetilde g}_{tk}} 
-{1\o 2} {\widetilde g}^{tk}\cancelto{0}{\p_k {\widetilde g}_{tt}}
= H {a^2 v^2\o c^2}
\ee
\be
{\widetilde \Gamma}^i_{tt} 
={1\o 2} {\widetilde g}^{it} {\p_t {\widetilde g}_{tt}} 
+ {\widetilde g}^{ik} {\p_t {\widetilde g}_{tk}} 
-{1\o 2} {\widetilde g}^{ik}\cancelto{0}{\p_k {\widetilde g}_{tt}}
= - 2 {v\o c} H + {v^3\o c^3} a^2 H
\ee

Just as for an object at rest in the istropic frame (\ref{ge}), the force equation in the boosted frame for an object at rest is
\be
{d{\widetilde U}^\mu\o d{\widetilde t}} + {\widetilde \Gamma}^\mu_{tt}c^2 = 0
\ee
where the proper time is just the coordinate time ${\widetilde t}$, and instantaneously ${\widetilde U}^\mu = (c,0,0,0)$.

In the instantaneous rest frame of the body, any force will manifest in the spatial components of the acceleration, since
\be
{d\o d\tau} ({\widetilde g}_{\mu\nu} {\widetilde U}^\mu {\widetilde U}^\nu) = 0 = 2 {\widetilde g}_{\mu\nu} {\widetilde U}^\mu {d{\widetilde U}^\nu\o d\tau}
\quad\rightarrow\quad {d{\widetilde U}^\nu\o d\tau} = \l(0, {d{\widetilde U}^x\o d\tau},0,0\r)
\ee
This means that an acceleration exists in the rest frame of an accelerated body, and in that frame the acceleration 4-vector is purely spatial. 

When we consider $d{\widetilde U}^x / d\tau$ in the instantaneous rest frame, we find a non-zero force component along the x-direction:
\be\label{hd3}
-{d{\widetilde U}^x\o d{\widetilde t}} = -{\widetilde \Gamma}^x_{tt}c^2 \quad\rightarrow\quad
{d{\widetilde U}^x\o d{\widetilde t}} = -2 {v\o c}H + {v^3\o c^3}a^2 H
\ee
% NI 3July: possible footnote stating that transf 30 is formally a non-relativistic transf, and therefore implies terms quadratic and higher order in v should be neglected.
The form of the rest-frame force (\ref{hd3}) is the same as (\ref{hdu}). The acceleration measured on the body from the isotropic frame is the same force the body feels in its rest frame. This shows that this force is frame-independent. 

There are an infinity of reference frames in which Hubble drag arises as it does here, from the off-diagonal, 3-vector metric components ${\widetilde g}_{tk}$. There is only one frame in which it arises from isotropic metric components. Yet the frame-independence of the effect insures that it can be fairly called frame dragging.

\begin{comment}
There is also a rest-frame energy effect, even though for test particle there is no internal energy.
\be
{d{\widetilde U}^\mu\o d{\widetilde t}} = - a^2 v^2 H
\ee
which is equivalent to (\ref{hc}). As before, applied to galaxy clusters, it must be understood as a reduction of kinetic energy density in a volume of space.
\end{comment}

%%%%%%%%%%%%%%%%%%%%%%%%%%%%%%%%%%%%%%%%%%%%%
%%%%%%%%%%%%%%%%%%%%%%%%%%%%%%%%%%%%%%%%%%
\section{4. Inductive frame dragging}
%%%%%%%%%%%%%%%%%%%%%%%%%%%%%%%%%%%%%%%%%%%%%
%%%%%%%%%%%%%%%%%%%%%%%%%%%%%%%%%%%%%%%%%%

Gravitomagnetism, frame-dragging, and Lense-Thirring effect are synonymous terms for the same underlying effect. They arise from the influence of off-diagonal metric elements on the motion of bodies. These effects are seen in both linear approximations and exact solutions of the Einstein equations. We briefly review both.

The Kerr metric is an exact solution for a static, spherical spacetime with aziumthal symmetry around an object with mass $M$ and angular momentum $J$. The metric in Boyer-Lindquist coordinates can be approximated in a weak field limit:
\be
\label{kerr}
-c^2 d\tau^2 \simeq -\l( 1 - {2M\o r}\r) c^2dt^2 + \l( 1 + {2M\o r}\r)dr^2 + r^2(d\theta^2 + \sin^2\theta d\phi^2) - {4J\o r} \sin^2\theta d\phi cdt
\ee
Frame-dragging effects arise from the off-diagonal components $g_{t\phi}$, which have their source in the angular momentum of the central gravitating object.

It is also useful to consider gravito-magnetic effects as they emerge in the geodesic equation written to linear order in metric perturbations $h_{\mu\nu}$, where $g_{\mu\nu}\simeq \eta_{\mu\nu} + h_{\mu\nu}$. To linear order in test particle speed, the geodesic equation linear in the metric perturbations has the form\cite{wi}:
\be
\label{linmom}%7.26
{1\o m}{d p^i\o dt} \simeq
c^2 {E}^i_g + \epsilon_{ijk}c{v^j} B_g^k  
- {v^j} \p_t h_{ij} \quad ,\quad v\ll c
\ee
where $p^i$ are the spatial momentum components of a body in motion, $m$ is rest mass, and we have defined natural gravito-electric and -magnetic fields:
\be
\label{gravebfield}%7.25
E_g^i \equiv  \p_i h_{tt}/2 - \p_t h_{ti} /c 
\quad , \quad
B_g^i \equiv \epsilon_{ijk} \p_j h^{tk} 
\ee
The linear force equation (\ref{linmom}) carries a striking analogy to the Lorentz force of electromagnetism, but there are also non-Lorentz, tensor effects. The gravito-magnetic effect is apparent, and its potentials are the off-diagonal metric components $h_{tk}$. 

In the linear approximation, the off-diagonal components of (\ref{kerr}) in turn have their origin in static mass currents\cite{wi}:
\be
\label{mc}
\nabla^2 h_{tk} \simeq -{16\pi G\o c^4} T_{tk}
\ee
where $\nabla^2$ is the 3-space Laplacian and where the $T_{tk}$ are the time-space components of the energy-momentum sources. The units of $T_{tk}$ are momentum density or energy flux, and so they represent a mass current. Along with the transverse constraint that $\p_k h^{tk}=0$, (\ref{mc}) governs two of the six physical degrees of freedom of the linear gravitational field.\cite{wi} Those degrees of freedom are clearly elliptical, with sources in the mass currents.

Standard treatments of frame dragging utilize mass currents (\ref{mc}) originating in the rotation of astrophysical masses. There is a wide literature on this subject, but key works include \cite{trg}, \cite{ltg}, \cite{bc}, \cite{bc2}. See also \cite{pfr} and \cite{cfi1} for a review, and \cite{cfi2} for a measurement review. A cosmological metric has also been investigated as the boundary for a rotational geometry.\cite{kln},\cite{smd}. 

There is some association of frame dragging with accelerated masses, since rotating mass currents are sustained via centripetal acceleration. Therefore, the small literature on rectilinear frame dragging\cite{ge,lb,fz,pk,pfr,pfs} is focused on linear acceleration. Yet we see from (\ref{mc}) that acceleration is not required for a mass flux to exist. Rectilinear frame dragging for uniform motion is suggested here for the first time. 

Let us return to the connections in the geodesic equation (\ref{ge}). As seen in (\ref{c2}), the 3-space forces for a body in motion in the isotropic cosmological metric arise from:
\be
\label{force2}
2\Gamma^i_{tj} =  g^{ik} ( \underset{\text{Hubble drag}}{\p_t g_{jk}} + \underset{\text{gravito-magnetic force}}{[\cancel{\p_j g_{tk}} - \cancel{\p_k g_{tj}}]} ) 
\ee
The forces associated with the connections (\ref{force2}) are proportional to particle velocity, and we have identified the separate contributions of Hubble drag and gravito-magnetic forces.

On the other hand, for a homogeneous metric ${\widetilde g}_{\mu\nu}$ with off-diagonal components ${\widetilde g}_{tk}$, the 3-forces on a body at rest arise from:
\be
\label{force}
2{\widetilde \Gamma}^i_{tt} = {\widetilde g}^{it}  \p_t {\widetilde g}_{tt} 
+ \underset{\text{gravito-electric force}} {[\overset{\text{inductive}}{2{\widetilde g}^{ik}  \p_t {\widetilde g}_{tk}}- \overset{\text{Newtonian}}{ {\widetilde g}^{ik}  \cancel{\p_k {\widetilde g}_{tt}}} ]}
\ee
These forces exist for bodies at rest, and contain the limit of Newtonian gravity as well as a gravito-electric induction force.

The inductive force effects were considered early by Einstein\cite{ein}, and he coined the term ``inductive", although he anticipated the inductive effects would come from accelerated motion, not from the expanding spacetime. 

For the homogeneous cosmological metric, spatial derivatives vanish. The Newtonian piece of the gravito-electric field (\ref{force}) vanishes, as does the entire gravito-magnetic field (\ref{force2}). Hubble drag in the isotropic frame is accounted in the boosted frame by the inductive term in the gravito-electric field. Therefore, we call this effect ``inductive linear frame dragging".

There are some important distinctions from frame-dragging induced by rotating mass currents. One is that a gravito-magnetic force in (\ref{linmom}) acts normal to particle velocity, and so does no work. The forces of Hubble drag and the inductive gravito-electric effect, by contrast, do work on a body, changing its energy.

Another distinction is the Hubble drag force, or the inductive dragging, occurs at constant velocity. It does not require acceleration, as in frame-dragging from rotation. Therefore we consider this the first proposed case of uniform-velocity frame dragging.

Any velocity transformation will introduce off-diagonal components into the metric, so there is some reasonable question as to whether such effects are real or are a coordinate artifact. Ref.~\cite{cfi1} suggested a scalar invariant for the existence of gravitomagnetism in terms of the Riemann tensor and the antisymmetric tensor:
\be
\label{ci}
I = \epsilon^{\alpha\beta\sigma\rho}R_{\sigma\rho\gamma\delta} R_{\alpha\beta\mu\nu} g^{\mu\gamma}g^{\nu\delta}
\ee
The quantity $I$ is zero for the metric (\ref{rwm}). However, the traditional scalars of the flat Robertson-Walker metric, the curvature scalar $R$ and the Kretschmann scalar $\mathscr{R}$, are non-zero:
\be
c^2 R = 6 \l( {\bm\dot a^2\over a^2} + {\bm\ddot a\o a} \r)
\quad,\quad c^4 \mathscr{R} = 12 \l( {\bm\dot a^4\over a^4} + {\bm\ddot a^2\o a^2} \r)
\ee

There can be no traditional gravito-magnetic invariant involving contractions of Riemann for uniform-velocity frame dragging in the cosmological metric because rectilinear uniform motion is coordinate-dependent.

This is unlike the traditional gravito-magnetic invariants, or the Kerr metric itself. While the off-diagnonal elements of the Kerr metric arise from mass in motion, the motion is compact, and cannot be transformed away with respect to the background stars. This is consistent with the understanding that the de Sitter precession arising from motion around a static mass source is fundamentally different than frame-dragging or gravitomagnetism.\cite{cfi1,cfi3} Therefore the metric arising from the closed mass flux of a rotating central object can be described in terms of an invariant composed of contractions of the Riemann tensor.

The compactness of the matter current not only allows an invariant, but also implies acceleration, and establishes the link between gravitomagnetism and acceleration. Acceleration is not necessary for an inductive rectilinear gravito-electric effect, but it will be coordinate dependent in the way of de Sitter precession.

This implies that there are dragging phenomena that are not contemplated in gravito-magnetic invariants of the Riemann tensor.

%%%%%%%%%%%%%%%%%%%%%%%%%%%%%%%%%%%%%%%%%%%%%%%%%%%%%
%%%%%%%%%%%%%%%%%%%%%%%%%%%%%%%%%%%%%%%%%%%%%%%%%%%%
\section{5. Local coupling to the cosmological metric}
%%%%%%%%%%%%%%%%%%%%%%%%%%%%%%%%%%%%%%%%%%%%%%%%%%%%%%%%
%%%%%%%%%%%%%%%%%%%%%%%%%%%%%%%%%%%%%%%%%%%%%%%%%%%%%%%%

The Hubble drag force in principle acts on bodies great and small, and so represents a local momentum transfer from matter to the gravitational field of the universe. This locality of momentum lost by the matter is not mirrored by a localization of momentum deposited into the field, because gravitational field energy-momentum cannot be localized. The freedom of a general coordinate transformation allows construction of a local inertial frame in which all gravitational forces vanish, and so local gravitational energy-momentum must vanish.\cite{mtw}

Yet gravitational forces do work locally on matter, as can be seen from the covariant divergence of the Einstein equations \cite{wbgem}:
\be
\nabla_\mu T^{\mu\nu} = g^{-1/2} \p_\mu (g^{1/2} T^{\mu\nu}) + \Gamma^\nu_{\mu\lambda} T^{\mu\lambda} =0
\ee
This is the expression of conservation of energy and momentum in a gravitational field. The last term represents the energy-momentum exchanged between the non-gravitational energy-momentum $T^{\mu\nu}$ and the gravitational field.

The energy-momentum of the gravitational field can be characterized by decomposing the metric into a Minkowski piece and a variable piece $f_{\mu\nu}$ that is not small:
\be
g_{\mu\nu} \equiv \eta_{\mu\nu} + f_{\mu\nu}
\ee
With this, the Einstein tensor can be decomposed into a term linear in $f_{\mu\nu}$ and terms of higher order:
\be
G_{\mu\nu} \equiv G^{(1)}_{\mu\nu} + G^{(2+)}_{\mu\nu}
\ee
Then the Einstein equations can be recast exactly:
\be
G^{(1)}_{\mu\nu} = {8\pi G}T_{\mu\nu} -G^{(2+)}_{\mu\nu}
\ee
This invites definition of the gravitational energy-momentum pseudotensor:
\be
\Theta_{\mu\nu} \equiv (8\pi G)^{-1}[G_{\mu\nu} - G^{(1)}_{\mu\nu}]
\ee
The pseudotensor $\Theta_{\mu\nu}$ has several properties that support its interpretation as the energy-momentum of the gravitational field \cite{wbgpt}. One is that the total energy-momentum of matter and the field is a constant:
\be
\label{tem}
\p_\nu [ \eta^{\nu\mu}\eta^{\lambda\kappa} ( T_{\mu\kappa} + \Theta_{\mu\kappa} ) ] \equiv \p_\nu ( T^{\nu\kappa} + \Theta^{\nu\kappa} )= 0
\ee

The pseudotensor $\Theta^{\nu\kappa}$ describes the gravitational field energy measured in any coordinate system. However, the pseudotensor is not a tensor, and is coordinate-dependent. It yields meaningful integrals of energy and momentum only if the fields asymptotically approach Minkowskian.\cite{wbgpt}

Consideration of the volume integral of (\ref{tem}) allows identification of the energy-momentum 4-vector that locally describes the combined gravitational and non-gravitational contributions at every spacetime point:
\begin{equation} %W7.6.9
\label{mom}
    P^\mu = \int  ( T^{\mu t} + \Theta^{\mu t} )d^3x
\end{equation}
where the $t$ index indicates the time component.

Since the field energy cannot be localized, the momentum absorbed by the gravitational field of the universe $\Theta^{j t}$ owing to rectilinear inductive frame dragging of matter $T^{j t}$ must be expressed as an integral of $\Theta^{j t}$ over a finite region, where roman indices indicate the 3 spatial coordinates. Therefore, from (\ref{mom}) the momentum lost by a body moving with velocity $v$ for a time $\Delta t$ would be absorbed into the field in a volume of size $\Delta x=v\Delta t$:
\be  
\label{mt}
\int T^{ti} d^3 x = -\int \Theta^{ti} d^3x = - v \int \Theta^{ti}\Delta t\ d^2x %LW: corrected 5July
\ee
%The equation above is for a test body, and so is the energy-momentum. It describes that the momentum lost by a body under Hubble drag is taken up into the energy-momentum of the gravitational field over a region large enough to contain the dynamics.

Cosmologically, $\Theta^{ti}=0$ for (\ref{rwm}), indicating that the gravitational field of the universe carries no net momentum. However, the cosmological gravitational field is integrated over all matter in the universe, with varying concentrations and in varying states of motion. The total gravitational field will include all these effects from source motion, including the effects in the field of local mass currents. However, the cosmological field will always naturally be described in the frame in which the net motion of the matter in the universe $T^{it}_U$ is zero, and the net motion of matter in the universe can always be determined from an integral over all space:
\begin{equation}
    \oint T_U^{it}d^3x \equiv \rho_U c^2 V^i
\end{equation}
The metric (\ref{rwm}) is then formulated in the frame for which $V^i=0$, composed as it is of all the disparate motion of all the matter in the universe.

Since matter can lose momentum into the gravitational field of the universe, it seems to imply that the gravitational field of the universe can be detected from dynamics. The Hubble constant can be measured by looking outward into space, at galactic redshift. It would appear that Hubble drag also allows such a determination looking inward at the motion of material bodies.

%%%%%%%%%%%%%%%%%%%%%%%%%%%%%%%%%%%%%%%%%%%%%%%%
%%%%%%%%%%%%%%%%%%%%%%%%%%%%%%%%%%%%%%%%%%%%%%%%%%
\section{6. Detectability of Hubble drag in the laboratory}
%%%%%%%%%%%%%%%%%%%%%%%%%%%%%%%%%%%%%%%%%%%%%%5
%%%%%%%%%%%%%%%%%%%%%%%%%%%%%%%%%%%%%%%%%%%%%%
Cosmological redshift is given in terms of the scale factor by the simple textbook formula
\begin{equation}
\label{cr}
    {\lambda_e\o a(t_e)} = {\lambda_r\o a(t_r)}
\end{equation}
where the subscripts indicate ``emission" and ``reception" of a photon, $t$ is the time, and $\lambda$ the wavelength. The Hubble expansion famously redshifts light, and the redshift increases with time. The galaxies can be considered beacons emitting light at known frequencies, and the cosmological redshift increases the separation between beacons at rest, and reddens their light. Since the momentum of a photon is inversely proportional to the wavelength, the reddening of (\ref{cr}) can be seen to correspond to the same momentum loss described by Hubble drag.

We have also seen in (\ref{hd}) that the geodesic equation describes an analogous loss of momentum for massive bodies, not just for light. In fact, the same term in the geodesic equation accounts for the effect on both timelike and null trajectories. Yet due to the ``running-awayness" of the momentum loss associated with Hubble expansion -- the passive draining of momentum, instead of losing it to work elsewhere in the system -- there is a misconception that Hubble drag is not a ``real" force. %Pinnochio is not a real boy!

In principle, an experiment could be conducted to detect the Hubble expansion within the bounded space of a laboratory, without looking at the distant galaxies or at the cosmic microwave background. In addition to the well-known effect on redshift that accrues over time as the universe expands, there is also an effect on the Doppler shift arising from Hubble drag. At intermediate cosmological redshifts, there is an interplay of these effects before the magnitude of cosmological redshift overwhelms the Doppler shifts in local velocity effects. Of course, Hubble drag has been modeled already statistically in galaxy dynamics in terms of ``peculiar velocities"\cite{P2}, but we are talking here of detecting it on a single moving body.

Let us consider a hypothetical laboratory deep in intergalactic space, far from local gravitational sources. 
%Alternatively, we can consider a system in local free fall near gravitational sources, but with the system much smaller than the scale of any tidal forces. 
%An object set in motion would immediately experience a deceleration from Hubble drag that would slow the object with time. 
Consider a test body carrying a beacon of known frequency, and receding along the line of sight. A beacon at rest in the frame of (\ref{rwm}) will have a redshift from the Hubble expansion, and this is the usual cosmological redshift. A beacon receding in the frame of (\ref{rwm}) will have an additional redshift from the Doppler effect. There is also a third redshift effect from the time dilation of the moving source that will always accompany its motion. Let us quantify this.

Let the radiation emitted from the beacon to be at a fixed frequency corresponding to some narrow spectral line. The period of the radiation in the rest frame of the moving body can be written as a differential $d{\widetilde t}$, which is an invariant:
\begin{equation}
    d{\widetilde t}^2 = \eta_{\mu\nu} d{\widetilde x}^\mu d{\widetilde x}^\nu = g_{\mu\nu} dx^\mu dx^\nu = g_{\mu\nu}{dx^\mu\o dt} {dx^\nu \o dt} dt^2
\end{equation}  %compare Weinberg 3.5.1
where $dt$ is the time in the frame of (\ref{rwm}).

In addition to this time dilation in curved space, the period $d t_o$ of the radiation received by an observer at rest in the cosmic isotropic frame will be shifted by the motion of the object along the line of sight:
\begin{equation}
\label{td}
    d t_o = (1+a v_r/c)d t = (1+a v_r/c) \l( g_{\mu\nu}{dx^\mu\o dt} {dx^\nu \o dt}\r)^{-1/2}d {\widetilde t}
\end{equation}  %compare Weinberg 2.2.2
where $v_r$ is the velocity of the source along the radial direction from the observer, $a$ is the scale factor, and $g_{\mu\nu}$ is given by (\ref{rwm}). For this time dilation expression, the Hubble timescale is much longer than the period of the radiation, so the term in $g_{\mu\nu}$ can be evaluated at a single instant. 

Let us now convert $(dx/dt) \rightarrow v$, and convert (\ref{td}) from period to frequency. Then $d t_o\rightarrow 1/f_o$ and $d{\widetilde t}\rightarrow 1/{\widetilde f}$, and the frequency shift of a beacon moving against the metric (\ref{rwm}) is
\begin{equation}
\label{dop}
    {f_o\o {\widetilde f}} = {(1-a^2 v^2/c^2)^{1/2}\o (1+av_r/c)} \quad \underset{v_r = v}{\rightarrow}\quad 
    \l( {1-av/c}\o {1+av/c} \r)^{1/2}
\end{equation}
where the last expression follows because the beacon is receding from the observer along the line of sight. Note that the previous 3 expressions relate the frequency of radiation seen in two different coordinate systems, but co-located at the same spacetime point. We have not yet integrated radiation from the moving source to a fixed observer.

Now let us evaluate the time derivative of $f_o$ in (\ref{dop}):
\begin{equation}
\label{tdv}
     {\bm\dot f_o\o {\widetilde f}} = -{f_o\o {\widetilde f}} { {(v\bm\dot a /c + a\bm\dot v/c)}\o (1-a^2v^2/c^2)}
\end{equation} %LW doublechecked
where dot notation $\bm\dot f_0$, $\bm\dot a$, and $\bm\dot v$ indicates time derivatives.

There are two effects on the time derivative of the Doppler shift (\ref{tdv}). One is from the time derivative of the scale factor, and represents the effect of the Hubble expansion carrying bodies apart. The other is from the time derivative of the beacon velocity, presumed to be receding along the line of sight. These two effects operate oppositely on the redshift.

The scale factor increases with time, so its time derivative is positive. Yet $\bm\dot v < 0$ according to (\ref{hdu}), due to the slowing of Hubble drag. Putting (\ref{hdu}) into (\ref{tdv}) yields:
\begin{equation}
    {\bm\dot f_o\o {\widetilde f}} = {{\bm\dot a v}\o {c}}{f_o\o {\widetilde f}} >0
\end{equation}
The Hubble drag effect dominates in the Doppler shift over the Hubble expansion effect, leading to a net positive time derivative of the Doppler-shifted frequency as measured at the beacon.

This makes sense physically. A beacon receding at velocity $v$ at time $t=0$ will have its maximum Doppler shift according to (\ref{dop}). As Hubble drag works to decelerate the beacon, the reduction in speed against the isotropic background leads to a decrease of the redshift at time $t >0$ compared to $t=0$. Although the Hubble expansion is always working to increase redshift, the deceleration of Hubble drag produces a stronger effect that leads to a net bluing of the Doppler redshift. As Hubble drag brings the object to rest over the age of the universe, the locally-measured Doppler and time-dilation redshift effects vanish. This is indicated schematically in Figure 1.
\begin{figure*}
\includegraphics[width=0.7\textwidth]{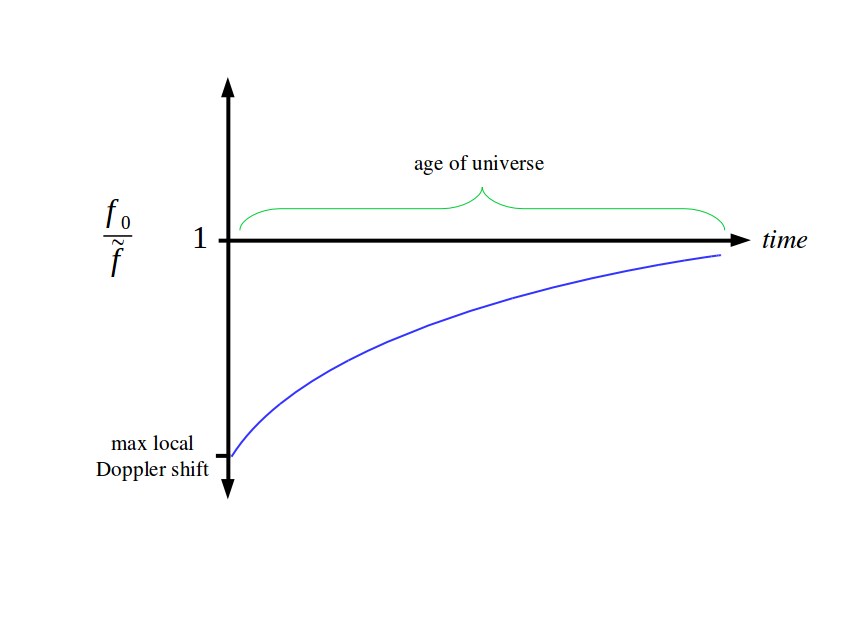}
\caption{Schematic diagram of the bluing of Doppler redshift accruing from Hubble drag. This diagram shows the time evolution of the locally-observed Doppler shift of a receding beacon. This frequency shift at the beacon includes effects from time dilation, recession, and Hubble expansion. Hubble drag will slow the velocity relative to the cosmic rest frame, manifesting as a decrease in locally-observed Doppler shift. \label{Fig1}}
\end{figure*}

In this way, by detecting the action of Hubble drag on the Doppler shift of a moving body, and the resulting frequency shift operating opposite to the cosmological redshift, it is in principle possible to determine cosmological parameters by measurements on material bodies, without looking at distant objects. 

The local detection of cosmic expansion through the time derivative of the redshift invites consideration of Mossbauer-type experiments. The Pound-Rebka experiment used this technique and obtained a frequency measurement of gravitational redshift to a part in $10^{15}$.\cite{pr} They used controlled Doppler shifts to cancel out the gravitational redshift over a 20-meter vertical drop.

The sensitivity of gravitational redshift detection can be compared with the necessary sensitivity of a Hubble drag experiment. In that case, the size of the effect (\ref{sz}) is the ratio of the dynamic timescale to the age of the universe. For a timescale of hours, the ratio is of order $10^{-14}$. The associated frequency shift is reduced by a factor of $v/c$, so that a beacon moving at 10\% the speed of light would show a shift detectable at the Pound-Rebka resolution. Configuring a free-fall experiment that eliminates all gravitational effects except Hubble expansion might be impractical in a terrestrial experiment.

Nonetheless, a redshift observation of this type would be described by the time integral of radiation received from the decelerating source:
\begin{equation}
\label{ds}
   \Delta f_0 \equiv \int_{t_e}^{t_r} \bm\dot f_o(t') dt' = {\widetilde f}
     \int_{t_e}^{t_r} {{\bm\dot a v(t')}\o {c}}
     \l( {1-av/c}\o {1+av/c} \r)^{1/2} dt'
\end{equation}
where $v(t)$ is the solution to (\ref{hdu}), which depends on the scale factor $a(t)$, which in turn depends on the cosmological energy densities. Therefore, the solution to (\ref{hdu}) is a well-defined, albeit complex, function of the standard cosmological model. The cosmology-dependent Hubble drag $v(t)$ is then integrated along the line of sight in (\ref{ds}) to yield the total frequency shift measured by an observer stationary in the frame of (\ref{rwm}). As Hubble drag slows the beacon over the age of the universe, its locally-measured Doppler shift vanishes as shown in Fig.1, and the redshift seen by a fixed observer merges with the average Hubble flow measured with standard candles at rest in the frame of (\ref{rwm}).

This shows the momentum transfer of Hubble drag is real and measurable, and in principle observable looking inward at a laboratory configuration large enough to observe a fast body for approximately an hour.

%%%%%%%%%%%%%%%%%%%%%%%%%%%%%%%%%%%%%%%%%%%%%
%%%%%%%%%%%%%%%%%%%%%%%%%%%%%%%%%%%%%%%%%%
\section{7. Conclusions}
%%%%%%%%%%%%%%%%%%%%%%%%%%%%%%%%%%%%%%%%%%%%%
%%%%%%%%%%%%%%%%%%%%%%%%%%%%%%%%%%%%%%%%%%
\begin{enumerate}

\item The Hubble drag force exists for motion with respect to the isotropic galactic free-fall frame because the isotropy of the cosmological metric picks out a preferred frame. 

\item The energy and momentum lost by Hubble drag are dissipated into the gravitational field of the universe on cosmological timescales.

\item The effect of Hubble drag on Doppler shift is larger and opposite to that of the Hubble expansion, leading to a bluing of Doppler-redshifted objects over the age of the universe.

\item Since Hubble drag always operates on test bodies, it implies the cosmological metric can be measured in principle through laboratory-scale dynamical experiments. If resolution similar to the Pound-Rebka experiment were attainable, Hubble drag should be detectable in free-fall Doppler shifts over an integration time of order 1000 seconds for bodies with $v/c \sim 0.1$.

\item These considerations show there is a channel for exchange of energy and momentum between a test body and the gravitational field of the universe. This is stated mathematically in (\ref{mt}).

\item The isotropic cosmological metric acquires off-diagonal components in boosted frames. In these frames, the Hubble drag force can be considered inductive rectilinear frame-dragging. It arises from the inductive part of the gravito-electric force, and constitutes a type of frame-dragging that is not covered by usual gravito-magnetic and frame-dragging invariants.

%\item Hubble drag should be considered a local gravitational force effect from the gravitational field of the universe. %This effect is dissipative, and so is suggestive that perhaps inertia, the resistance to acceleration, may also have a gravitational origin. The Hubble drag force is not large enough to explain inertia.

%\item An invariant in first derivatives of the metric captures the frame-independence of Hubble drag and rectilinear frame-dragging.

\end{enumerate}

%%%%%%%%%%%%%%%%%%%%%%%%%%%%%%%%%%%%%%%%%%
\vspace{6pt}

%%%%%%%%%%%%%%%%%%%%%%%%%%%%%%%%%%%%%%%%%%
\section{8. Acknowledgements}
%%%%%%%%%%%%%%%%%%%%%%%%%%%%%%%%%%%%%%%%%%
This research was funded by DARPA DSO under award number D19AC00020.

Yaroslav Balytski evaluated the gravito-magnetic scalar with tensor algebra software.}

%%%%%%%%%%%%%%%%%%%%%%%%%%%%%%%%%%%%%%%%%%
%\end{paracol}
%\reftitle{References}
%%%%%%%%%%%%%%%%%%%%%%%%%%%%%%%%%%%%%%%%%%%%%
%%%%%%%%%%%%%%%%%%%%%%%%%%%%%%%%%%%%%%%%%%
\section{9. References}
%%%%%%%%%%%%%%%%%%%%%%%%%%%%%%%%%%%%%%%%
%%%%%%%%%%%%%%%%%%%%%%%%%%%%%%%%%%%%%%%%%


\begin{thebibliography}{}

\bibitem{wbg} S. Weinberg, {\it Gravitation and Cosmology}, John Wiley \& Sons: New York (1972); see Section 3.2.

\bibitem{pw} E. Poisson and C. Will, \textit{Gravity}, Cambridge University Press: Cambridge (2014). see section 5.2.5

\bibitem{mtw}C. Misner, K. Thorne, \& J. Wheeler, {\it Gravitation}, Freeman: San Francisco, (1972). see Section 20.4


\bibitem{fth} J. A. Frieman, et al., Dark energy and the accelerating universe, {\it Ann.Rev.Astron.Astrophys.}, {\bf 46}, 385 (2008)

\bibitem{crl} S.M. Carroll, {\it Spacetime and Geometry}, Addison-Wesley: San Francisco (2004). See section 8.5

\bibitem{isl} J.N. Islam, {\it An Introduction to Mathematical Cosmology}, Cambridge University Press: Cambridge (2002). See section 2.4.

\bibitem{P1} P.J.E. Peebles, {\it Physical Cosmology}, Princeton University Press: Princeton (1971)

\bibitem{P2} P.J.E. Peebles, The peculiar velocity field in the local supercluster, {\it Ap.J.}, {\bf 205}, 318 (1976)

\bibitem{pck} J.A. Peacock, {\it Cosmological Physics}, Cambridge University Press: Cambridge (1999). See chapter 15 

\bibitem{ldr} E.V. Linder, Cosmic growth history and expansion history, {\it Phys.Rev.D}, {\bf 72}, 043529 (2005)

\bibitem{wbg2} Ref.~\cite{wbg}, Section 14.1

\bibitem{ge} O. Gron \& E. Eriksen, Translational inertial dragging, {\it Gen.Rel.Grav.}, {\bf 21}, 105 (1989).

\bibitem{lb} D. Lynden-Bell et al., On accelerated inertial frames in gravity and electromagnetism, {\it Ann.Phys.}, {\bf 271}, 1 (1999)

\bibitem{fz} H. Farhoosh and L. Zimmerman, Killing horizons and dragging of the inertial frame about a uniformly accelerating particle, {\it Phys.Rev. D}, {\bf 21}, 317 (1980)

\bibitem{pk} H. Pfister \& M. King, {\it Inertia and Gravitation}, Lecture Notes in Physics 897, Springer: Heidelberg (2015).

\bibitem{pfr} H. Pfister, Dragging effects near rotating bodies and in cosmological models, in {\it Mach's Principle: From Newton's Bucket to Quantum Gravity}, J. Barbour \& H. Pfister, eds., Einstein Studies Volume 6, Birkhauser: Boston (1995)

\bibitem{pfs} H Pfister et al, A model for linear dragging, {\it Class. Quantum Grav.}, {\bf 22}, 4743 (2005)

\bibitem{wi} L. Williams \& N. Inan, Maxwellian mirages in general relativity, {\it New J.Phys.} (2021)

\bibitem{trg} H. Thirring, On the effect of rotating distant masses in Einstein’s theory of gravitation, {\it Zeit.Phys.}, {\bf 19}, 33-39 (1918)

\bibitem{ltg} J. Lense \& H. Thirring, On the influence of the proper rotation of a central body on the motion of the planets and the moon, according to Einstein’s theory of gravitation, {\it Zeit.Phys.}, {\bf 19}, 156-163 (1918)

\bibitem{bc} D.R. Brill \& J.M. Cohen, Rotating masses and their effect on inertial frames, {\it Phys.Rev}, {\bf 143}, 1011-1015 (1966)

\bibitem{bc2} D.R. Brill \& J.M. Cohen, Further examples of Machian effects of rotating bodies in general relativity, {\it Nuov.Cim.}, {\bf 56B}, 209-219 (1968)

\bibitem{kln} C. Klein, Rotational perturbations and frame dragging in a Friedmann universe, {\it Class.Quant.Grav.}, {\bf 10}, 1619 (1993)

\bibitem{smd} C. Schmid, Mach’s principle: Exact frame-dragging via gravitomagnetism in perturbed Friedmann-Robertson-Walker universes with K=(±1,0), {\it Phys.Rev.D}, {\it 79}, 064007 (2009)

\bibitem{cfi1} I. Ciufolini, Dragging of inertial frames, gravitomagnetism, and Mach's Principle, in {\it Mach's Principle: From Newton's Bucket to Quantum Gravity}, J. Barbour \& H. Pfister, eds., Einstein Studies Volume 6, Birkhauser: Boston (1995)

\bibitem{cfi2} I. Ciufolini, Dragging of inertial frames, {\it Nature}, {\bf 449}, 41-47 (2007)

\bibitem{cfi3} I. Ciufolini \& R. Matzner, Non-Riemannian theories of gravity and lunar satellite laser ranging, {\it Int.J.Mod.Phys. A}, {\bf 7}, 843 (1991)

\bibitem{ein} A. Einstein, {\it The Meaning of Relativity, Four Lectures Delivered at Princeton University, May 1921}, Princeton University Press: Princeton (1923). See Lecture IV, equation (118) and subsequent discussion.

\bibitem{wbgem} See Ref.~\cite{wbg}, 5.3.3.

\bibitem{wbgpt} See Ref.~\cite{wbg}, section 7.6.

\bibitem{pr} R.V. Pound \& G.A. Rebka, Jr., Gravitational red-shift in nuclear resonance,
{\it Phys.Rev.Lett.}, {\bf 3}, 439 (1959)

\end{thebibliography}
\end{document}